On the temperature of the moving rod

M. Kozlowski[1], J. Marciak – Kozlowska[2], M. Pelc[3]

[1] Institute of the Experimental Physics, Warsaw University, Warsaw, Poland
[2] Institute of Electron Technology, Warsaw, Poland
[3] Institute of Physics, Maria Curie – Sklodowska University, Lublin, Poland

Abstract

In this paper the Lorentz transformation for temperature is derived and discussed. Considering the hyperbolic heat transport equation it is shown that T( prim) = gamma T, where gamma is the Lorentz factor, *T* denotes the temperature in the rest system and T(prim) is the temperature in the moving system (rod).

1. Introduction

The relativistic formulation of thermodynamics was taken up already by Einstein himself and by several other physicist, notably Planck and von Laue. The principal result at this stage was the following

$$T = T^O \sqrt{1 - \frac{v^2}{c^2}} \quad (1)$$

In this equation T is the absolute temperature in a system which is moving with a velocity $\upsilon$ with respect to the rest system (which is indicated by a superscript 0)

In the paper by Ott [1] the traditional formulation (1) was questioned. Instead of the transformation (1) for the temperature T, Ott requires

$$T = \frac{T^O}{\sqrt{1 - \frac{v^2}{c^2}}} \quad (2)$$

In this paper considering hyperbolic heat transport equation the temperature transformation equation will be obtained. It will be shown that the Ott formulation is valid Lorentz transformation for temperatures.

2. The model equation

As was shown in monograph [2] the master equation for the heat transport induced by ultra-short laser pulses can be written as:

$$\vec{q} + \tau \frac{\partial \vec{q}}{\partial t} = -\kappa \frac{\partial T}{\partial x} \quad (3)$$

$$\frac{\partial \vec{q}}{\partial x} + c_V \frac{\partial T}{\partial t} = 0 \quad (4)$$

In Eq.(3) $\vec{q}$ is the heat current, T denotes temperature, τ is the relaxation time and κ is the heat conduction coefficient. In Eq.(4) $c_V$ is the specific heat at constant volume. For quantum limit of the heat transport τ is equal

$$\tau = \frac{\hbar}{m\upsilon^2} \quad (5)$$

where m is the mass of heat carriers and $\upsilon = \alpha c$. The constant $\alpha_i$ is the coupling constant, $\alpha_1 = 1/137$ for electromagnetic interaction and $\alpha_2 = 0.15$ for strong interaction, c is the vacuum light speed.

From Eqs (1) and (2) hyperbolic diffusion equation for temperature can be derived

$$\tau \frac{\partial^2 T}{\partial t^2} + \frac{\partial T}{\partial t} = D \frac{\partial^2 T}{\partial x^2},\tag{6}$$

where thermal diffusion coefficient, D,

$$D = \frac{\kappa}{c_V}.$$

The basic principle of the special relativity theory can be stated as:

All physics laws look the same in all inertial references frames. Equation (4) is the conservation of thermal energy. Let us consider two infinite rods K and K′, where K′ is moving with velocity v parallel to axis X. From the relativity principle we obtain

$$\frac{\partial \vec{q}}{\partial x} + c_V \frac{\partial T}{\partial t} = 0 \quad \text{in frame K}$$

$$\frac{\partial \vec{q}'}{\partial x'} + c_V \frac{\partial T'}{\partial t'} = 0 \quad \text{in frame K'} \tag{7}$$

As can be easily shown the $\vec{q}$, $c_V T$ and $\vec{q}\,'$, $c_V T'$ are transformed according to Lorentz transformation:

$$\begin{aligned}
q'_{x'} &= \gamma(q_x - v c_V T) \\
q'_{y'} &= q_y \\
q'_{z'} &= q_z \\
c_V T' &= \gamma\left[c_V T - \frac{v}{c^2} q_x\right]
\end{aligned}\tag{8}$$

and

$$\begin{aligned}
q_x &= \gamma(q'_{x'} + v c_V T') \\
q_y &= q'_{y'} \\
q_z &= q'_{z'} \\
c_V T &= \gamma\left[c_V T' + \frac{v}{c^2} q'_{x'}\right]
\end{aligned}\tag{9}$$

The current $q_x$ and temperature T form the four vectors.

The space time interval Δ,

$$\Delta = q_x^2 - c^2 c_V T^2 = q'^2_{x'} - c^2 c_V T'^2 \tag{10}$$

is invariant under the Lorentz transformations (8) and (9).

Let us consider the case where there is no heat current in rod K, i.e. when the rod has the temperature T = const and $\nabla T = 0$. In that case $q_x = 0$ and from formulae (8) we obtain

$$q'_x = -\gamma \upsilon c_V T$$
$$c_V T' = \gamma c_V T; \quad T' = \gamma T$$
$$T' > T \quad \text{for} \quad \gamma > 1 \tag{11}$$

Formula (11) is in agreement with Ott result. As the result we obtain that temperatures of the rods are different. And the moving rod observes the greater temperature T′ = γT.

From Eq.(10) we conclude that

$$q'^2_{x'} = c^2 c_V \left(T'^2 - T^2\right) \tag{12}$$

and $q^2_{x'} > 0, q_x' > 0$. It means that the heat current is directed parallel to the moving of the rod K in the reference frame K′.

References

[1] H. Ott, Z. Phys. 175 (1963) 70

[2] M. Kozlowski, J. Marciak – Kozlowska, Thermal processes using attosecond laser pulses, Springer 2006.